\documentclass[fleqn,12pt,twoside]{article}
\usepackage[headings]{espcrc1}

\readRCS
$Id: espcrc1.tex,v 1.2 2004/02/24 11:22:11 spepping Exp $
\usepackage{graphicx}
\usepackage[figuresright]{rotating}

\newcommand{\AmS}{{\protect\the\textfont2
  A\kern-.1667em\lower.5ex\hbox{M}\kern-.125emS}}

\title{Elliptic Flow at Finite Shear Viscosity in a Kinetic Approach at RHIC}

\author{V. Greco\address[MCSD]{Dipartimento di Fisica e Astronomia, Universit\'a di Catania, Via S.Sofia 64, 95125 Catania, Italy}\address[INFN]{INFN-LNS, Laboratori Nazionali del Sud, Via S.Sofia 62, 95125 Catania, Italy},
        M. Colonna\addressmark[INFN],
        M. Di Toro\addressmark[MCSD]\addressmark[INFN]
        and
        G.Ferini\addressmark[MCSD]}
       
\runtitle{Elliptic Flow at Finite Shear Viscosity in a Kinetic Approach at RHIC}
\runauthor{V. Greco}

\begin{document}

\maketitle

\begin{abstract}
Within a covariant parton cascade, we discuss the impact of both finite shear viscosity $\eta$ and 
freeze-out dynamics on the elliptic flow generated at RHIC.
We find that the enhancement of $\eta/s$ in the cross-over region of the QGP
phase transition cannot be neglected in order to extract the information from the QGP phase. 
We also point out that the elliptic flow $v_2(p_T)$ for a fluid at $\eta/s \sim 0.1-0.2$ is consistent
with the one needed by quark number scaling drawing a nice consistency between the nearly
perfect fluid property of QGP and the coalescence process.    
\end{abstract}

\section{Introduction}

The measure of the elliptic flow, $v_2(p_T)$, in the ultra-relativistic heavy ion collisions at the
Relativistic Heavy Ion Collider (RHIC) has revealed that the so-called quark-gluon plasma (QGP) is
an almost perfect fluid . However several approaches indicate that even a small finite value of the shear
viscosity to entropy density ratio $\eta/s \sim 0.1-0.2$ affect significantly the strength of $v_2(p_t)$ 
\cite{Romatschke:2007mq,Song:2008si}. Hence viscous corrections to ideal hydrodynamics are indeed large and causality and stability problems present in first order relativistic Navier Stokes hydrodynamics cannot be avoided \cite{Romatschke:2009im,Huovinen:2008te}. Second-order Israel-Stewart approach has been developed to simulate the RHIC collision providing a first estimate of the $\eta/s$ \cite{Romatschke:2009im}. Such an approach,
apart from the limitation to 2+1D simulations, has the more fundamental problem that it is based on a gradient expansion at second order that is not complete and that anyway cannot be sufficient to describe correctly the dynamics of a fluid with large $\eta/s$ as the one in the hadronic phase 
\cite{Romatschke:2009im,Huovinen:2008te}.

We have developed a covariant kinetic approach that is able to simulate a fluid 
at finite $\eta/s$ by mean of a local renormalization of the cross section, $\sigma\cdot \eta/s=\langle p \rangle/{15\, \rho}$,  similarly to \cite{Huovinen:2008te},   
see Ref.\cite{Ferini:2008he,Greco:2008fs} for more details.
This has the advantage to be a 3+1D approach not based on a gradient expansion that is valid also
for large viscosity and for out of equilibrium momentum distribution allowing a reliable
description also of the intermediate $p_T$ range where the important properties of quark number scaling
(QNS) of $v_2(p_t)$ have been observed \cite{Annual}.

\section{Impact of $\eta/s$ increase in the cross-over region}

We focus on the $Au+Au$ collisions at $\sqrt{s}=200$ AGeV employing standard Glauber initial conditions in r-space, a Boltzmann equilibrium distribution in momentum space for partons at $p_T<2$ GeV and a minijet
distribution at higher $p_T$. 

\begin{figure}[htb]
\includegraphics[scale=0.32]{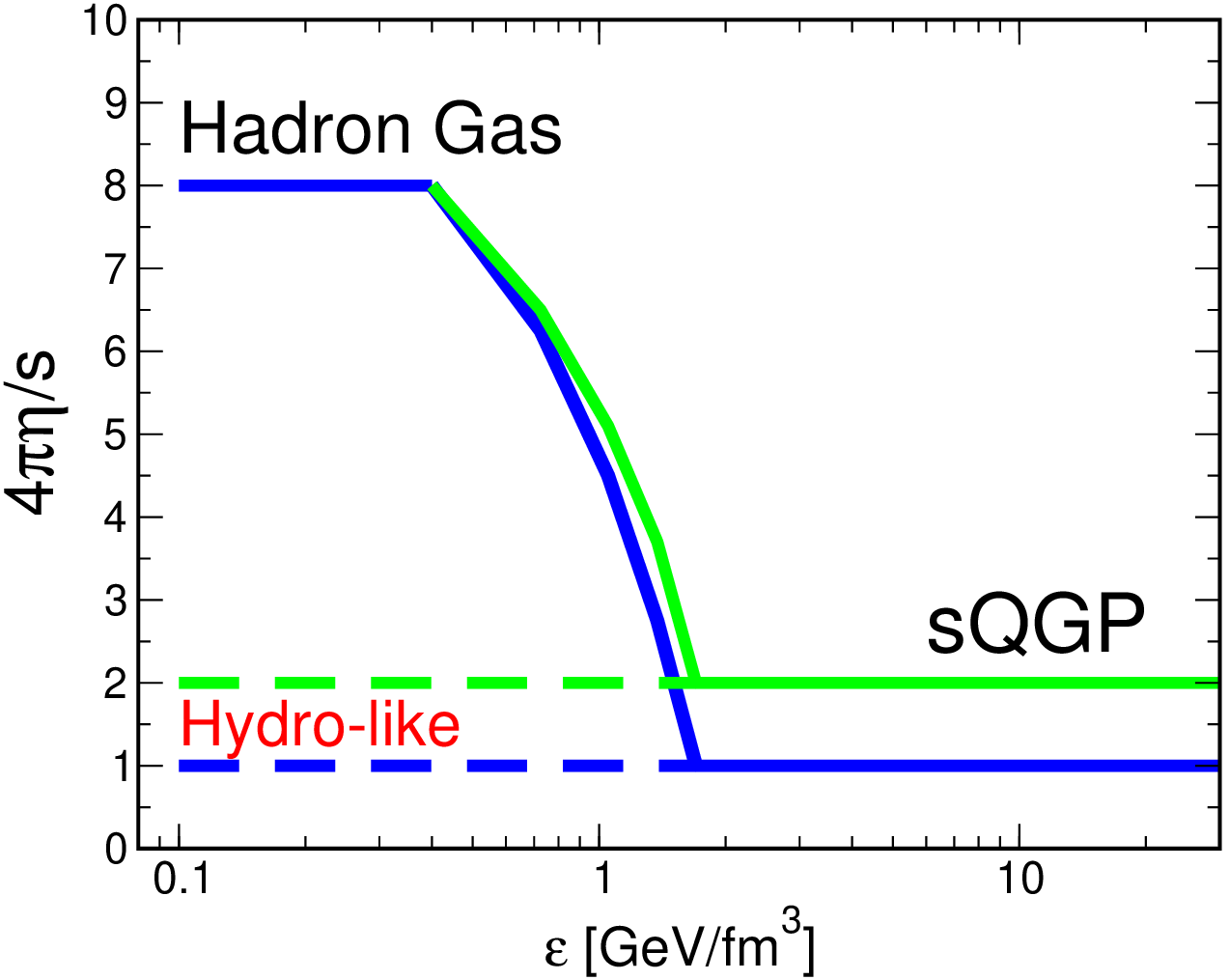}
\hspace{\fill}
\includegraphics[scale=0.37]{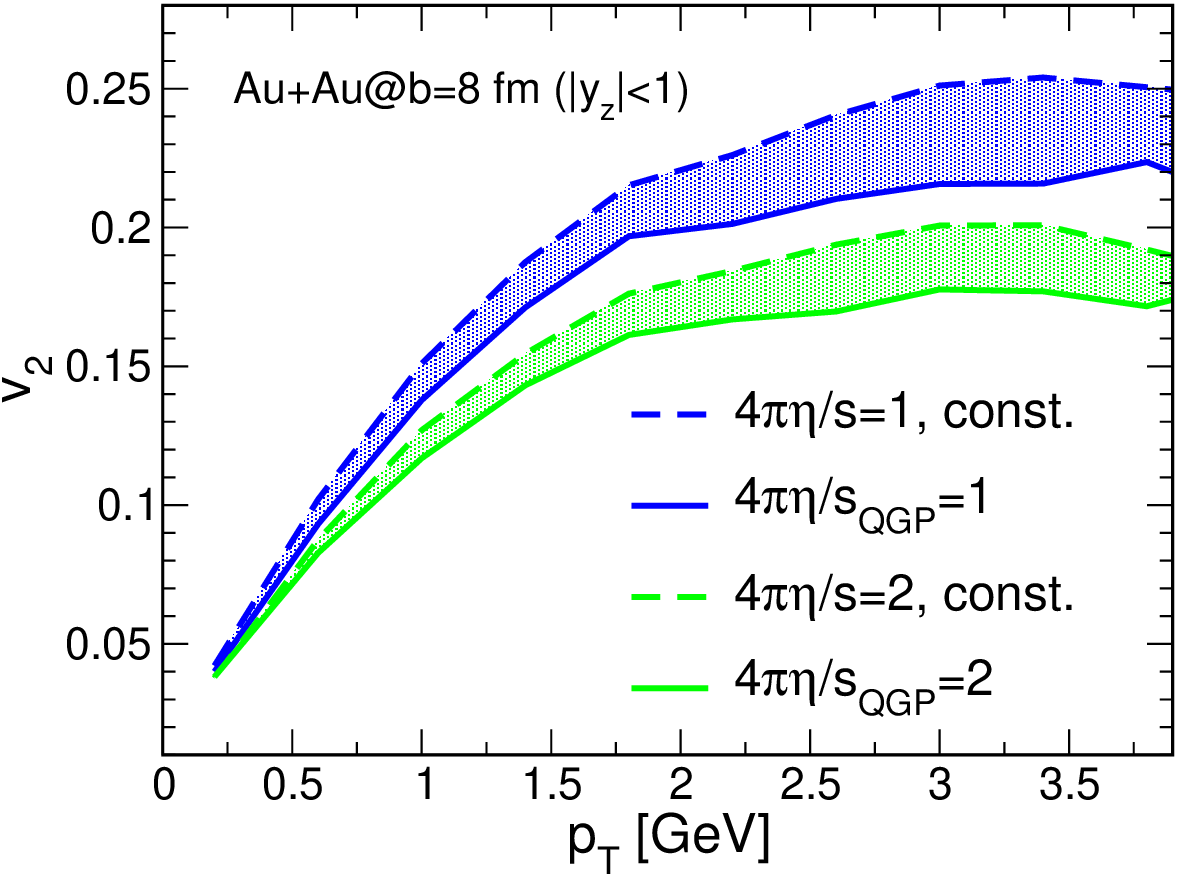}
\caption{Left: $4\pi\eta/s$ vs energy density $\epsilon$ for various simulations perfomed; Right: Elliptic flow
vs. $p_T$ for different cases; solid lines refer to calculation with $\eta/s$ increasing at 
$\epsilon<\epsilon_0$.}
\label{fig:hadro}
\end{figure}

A first objective is to evaluate the importance of the increase of the $\eta/s$ of the matter in the
cross-over transition and in the hadronic phase \cite{Demir:2009qi}. This is of particular relevance because most of the work
done till now to evaluate $\eta/s$ has been done in the viscous hydrodynamics framework keeping the 
$\eta/s$ constant during the entire evolution of the hadronic phase \cite{Romatschke:2007mq,Song:2008si}. 
As also mentioned in Ref.\cite{Romatschke:2009im}, it is desirable to take into account the evolution
of the $\eta/s$ inside and below the QCD phase transition. We have realized this imposing an increase
of the $\eta/s$ as a function of the local energy density, as shown in Fig.\ref{fig:hadro}.  
While in hydrodynamics the $\eta/s$ is kept constant during the entire evolution of the system (dashed
lines) in our calculation it increases when the cross-over region starts.  
The impact of such increase on the $v_2(p_T)$ is shown on the rigth side of Fig.\ref{fig:hadro}.
We see that even if most of the $v_2(p_T)$ is built up during the pure QGP phase, the cross-over
region can still produce a damping of the elliptic flow. 
Such a finding is similar to Ref.\cite{Hirano:2005xf}, but here
it is entwined in the context of QGP finite $\eta/s$ showing the relevance for its evaluation.  
From Fig.\ref{fig:hadro} we can deduce that neglecting the expected
increase of $\eta/s$ across the transition can introduce a sistematic error of the order of $40-50\%$.

\section{Scalings of $v_2$}

\begin{figure}[htb]
\includegraphics[scale=0.358]{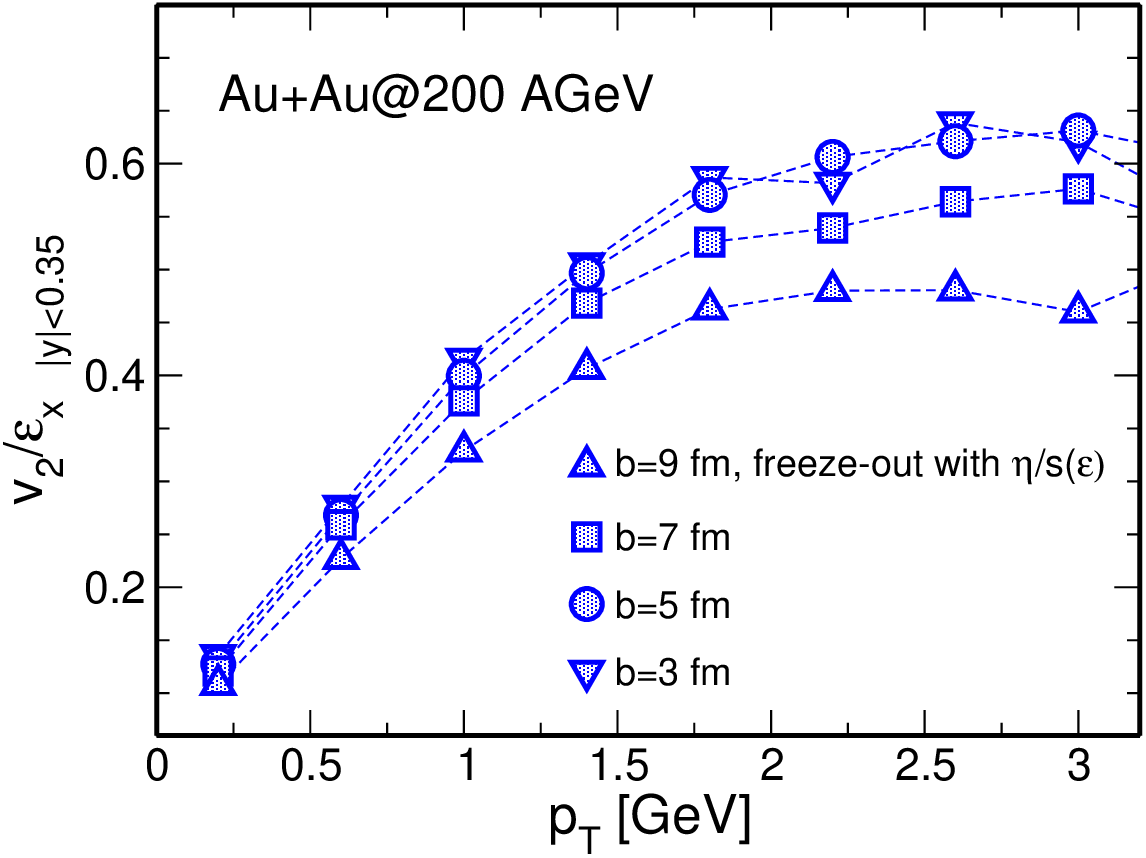}
\hspace{\fill}
\includegraphics[scale=0.34]{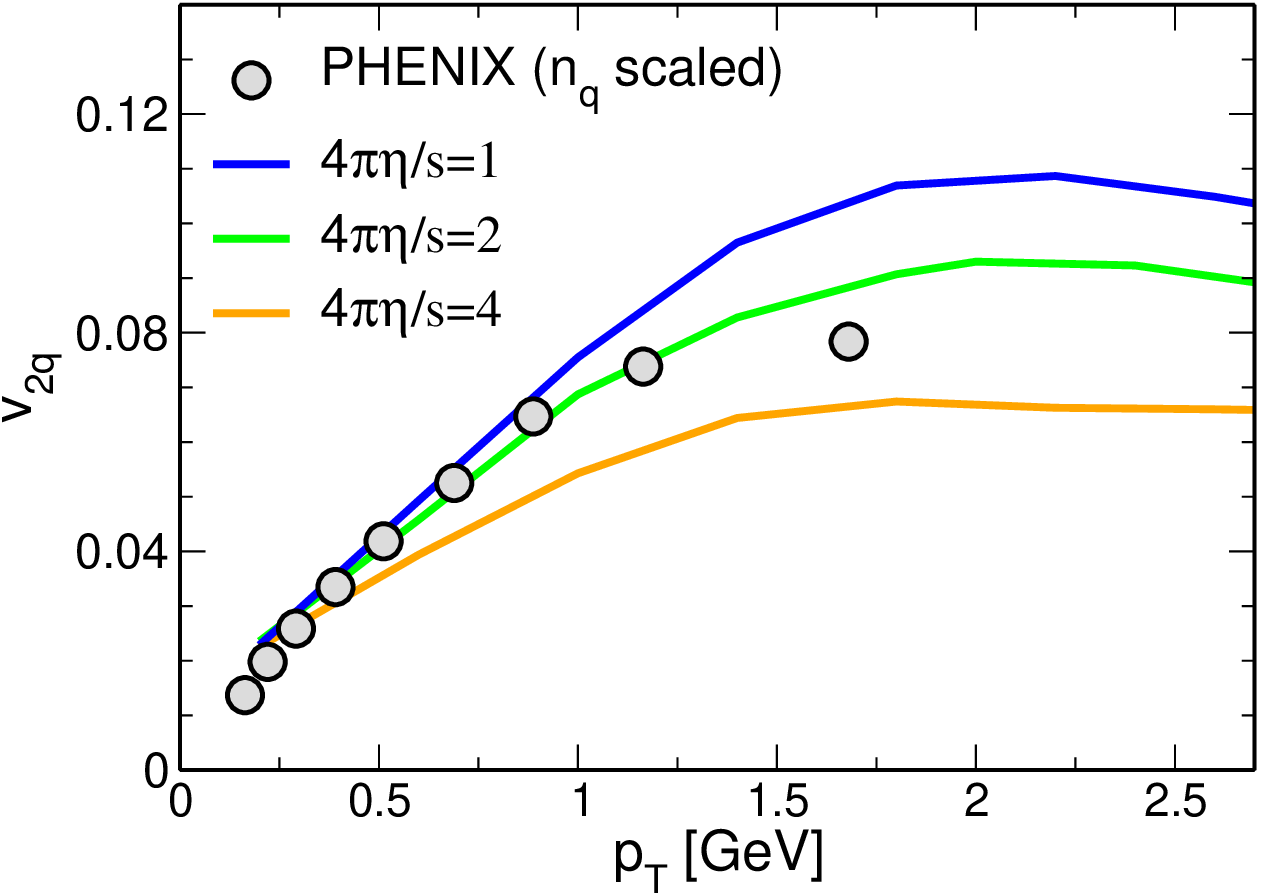}
\caption{Left: $v_2(p_T)$ over eccentricity for different impact parameters. Right: 
Elliptic flow vs. $p_T$ for three different value of $\eta/s_{QGP}$ at b=5fm. Circles are
data for the corresponding centrality \cite{Adare:2006ti} rescaled according to the $n_q$-scaling.}
\label{fig:v2pt}
\end{figure}

An energy density dependent $\eta/s$ represents also a way to realize a smooth kinetic freeze-out (f.o.)
of the system. 
In Ref.\cite{Greco:2008fs}, Fig.\ref{fig:v2pt} (left) here, we have shown that it is indeed the f.o.
mainly responsible for the observed breaking of the scaling of $v_2(p_T)$ with the initial space 
eccentricity $\epsilon_x$ \cite{Adare:2006ti,:2008ed}, at variance with ideal hydrodynamics prediction.

We notice the effect of viscosity increases with $p_T$ and is larger in the intermediate-$p_T$ region
see Fig.\ref{fig:v2pt}.
Moreover when both a suitable f.o. condition and a finite $\eta/s$ are taken into account
for the description of the fireball evolution, not only the breaking of the $v_2/\epsilon_x$ scaling
is reproduced along with the persistent $v_2(p_T)/<v_2>$ one \cite{:2008ed}, but also the shape of $v_2(p_T)$
is consistent with the one expected from QNS \cite{Annual}. In Fig.\ref{fig:v2pt} (right)
$v_2(p_T)$ at partonic level for different $\eta/s$ is shown together with the data from PHENIX, 
rescaled by the number of quarks that in the QNS scenario should correspond to quark one. We can see
that a $\eta/s\sim 0.15-0.2$ is consistent with $n_q$ scaling. Quantitatively a comparison with
experiments needs the inclusion of an equation of state with phase transition
that in the context of hydrodynamics has been shown to reduce significantly the $v_2$ \cite{Song:2008si} 
and threfore it can be envisaged to shift the agreement with the data down to $\eta/s \sim 0.1$.   
The inclusion of a mean field dynamics with phase transition is under investigation \cite{Plumari}.

\end{document}